# Strategies to integrate multi-omics data for patient survival prediction

Lana X Garmire
Department of Computational Medicine and Bioinformatics
University of Michigan
Ann Arbor, USA
lgarmire@med.umich.edu

Genomics, especially multi-omics, has made precision medicine feasible. The completion and publicly accessible multi-omics resource with clinical outcome, such as The Cancer Genome Atlas (TCGA) is a great test bed for developing computational methods that integrate multi-omics data to predict patient cancer phenotypes. We have been utilizing TCGA multi-omics data to predict cancer patient survival, using a variety of approaches, including using prior-biological knowledge (such as pathways), and more recently, deep-learning methods. Over time, we have developed methods such as Cox-nnet, DeepProg, and two-stage Cox-nnet, to address the challenges due to multi-omics and multi-modality. Despite the limited sample size (hundreds to thousands) in the training datasets as well as the heterogeneity nature of human populations, these methods have shown significance and robustness at predicting patient survival in independent population cohorts. In the following, we would describe in detail these methodologies, the modeling results, and important biological insights revealed by these methods.

**Survival prediction using biological functions (pathways) as features**

Conventionally, Cox-PH models are used to fit or predict patient survival. When genomics data (eg. transcriptomics) are used as the input features, certain feature selection or regularization methods (ridge, lasso, or elastic net) are used to help reduce the number of input features used to build multi-variate Cox-PH models. Many genes associated with the same biological functions are correlated. As a result, often times only a subset of genomics features is selected by the final model as potential biomarkers. Such approach has several potential issues for translational applications. One is that some features may not have clear or known biological functions, hampering them to be approved by FDA as biomarkers; another issue is that it may be difficult to ping-point the biological mechanism involved in patient survival.

To address these issues, we have utilized a personalized, pathway-level prognosis modeling approach, and applied it to predict breast cancer patient survival in several omics platforms(Al-Akwaa et al. 2018; S. Huang et al. 2014, 2016; Fang et al. 2020). The idea is to utilize Pathifier (Drier, Sheffer, and Domany 2013), an unsupervised dimension reduction method, to summary all genes' values into a single value that represents the pathway, called Pathway Dysregulation Score (PDS), among each patient. One can then use pathway PDS as input features to replace the original genomics features, before the conventional Cox-PH based prognosis prediction. Pathifer is based on the method of "Principle Curve" (Hastie and Stuetzle 1989), where PDS for a given pathway is the distance from the starting point of the principle curve to another projected point on the curve from the patient data point in the hyper-space. We applied gene expression microarray datasets of breast cancers to predict patient survival in several independent testing cohorts, and obtained significant difference in patient survival risk groups (S. Huang et al. 2014). Worth mentioning, the genomics to biological function (pathway) based approach is also applicable to disease diagnosis. We built a pathway based breast cancer diagnosis model using metabolomics data, and validated it with very high accuracy (AUC=0.99) in TCGA RNA-Seq data(S. Huang et al. 2016)

**Multi-modal data-based survival prediction using Cox-nnet models**

Rather than relying on conventional Cox-PH model, we sought to renovate survival prediction with neural-network (NN) models. Towards this we have developed Cox-nnet, a model where we add a fully connected NN to replace the linear model portion in Cox-PH(Ching, Zhu, and Garmire 2018). With a relatively simple three-layer Cox-nnet model, we tested on 10 different cancers in TCGA RNA-Seq data, and Cox-nnet achieved the same or better predictive accuracy compared to Cox-PH (with LASSO, ridge, and mimimax concave penalty), Random Forests Survival and CoxBoost methods. The hidden nodes Cox-nnet are surrogate features with prognosis predictive values, and it provides an alternative approach for survival-sensitive dimension reduction. Moreover, Cox-nnet demonstrates richer biological information, at both the pathway and gene levels.

We recently applied Cox-nnet to integrate multi-modal data types, exemplified by histopathology data and gene expression RNA-Seq data(Garmire et al. 2020). Compared to genomics data, pathological images are easily accessible data, however much more difficult to interpret. We first extended to predict patient survival using hepatocellular carcinoma (HCC) pathological images in TCGA. Again Cox-nnet based imaging predictions were more robust and accurate than Cox-PH model. We next used a two-stage Cox-nnet complex model, where the first stage Cox-nnet models were built on pathological image and RNA-Seq data respectively, we then combined the hidden layer nodes as the input for the 2nd-stage Cox-nnet model, yelling much more accurate prediction (C-index 0.77) than PAGE-Net, another complex, time-consuming Convolutional neural network (CNN) model, which had a median C-index of 0.68. The combined model was more predictive than models using either data type alone. There was modest correlation between the imaging and RNA-Seq features, where genes correlated with top imaging features were enriched in carcinogenesis pathways or with known association with survival of HCC patients and/or morphology of liver tissue.

**Multi-omics data based stratified survival risk prediction using DeepProg models**

Parallel to the strategy of modeling patient survival using Cox-nnet built on the assumptions of proportional hazard ratios and regressions, we also sought for an alternative approach that predicted patient prognosis risks (rather than survival time) using multi-omics data. This resulted in DeepProg method (Chaudhary et al. 2018). In this method, different omics data types are stacked together as inputs for an auto-encoder of five layers, upon fitting the hidden nodes in the bottleneck layer (3rd layer) are individually fit with Cox-PH model. The hidden features that are associated with survival are then used as the input for clustering. The best clusters are selected, and the patients are labeled into classes. Then classification model (eg SVM) is constructed using these labels and applied to predict other validation cohorts. We used RNA sequencing (RNA-Seq), miRNA sequencing (miRNA-Seq), and methylation data from hepatocellular carcinoma (HCC) patients in TCGA as the input of DeepProg, which yielded two optimal subgroups of patients with significant survival differences (C-index 0.68 and log-ranked p-values 7.13e-6). This multi-omics model was validated on five external datasets of various omics types with C-index between 0.67-0.82. Moreover, we found that the more aggressive subtype was associated with frequent TP53 inactivation mutations, as well as activated signaling pathways such as Wnt and Akt signaling pathways, whereas the patients with better survival have mostly metabolomic dysfunctions.

We recently upgraded DeepProg with better and performance, first by fitting each omics data type with a separate auto-encoder model (Poirion, Chaudhary, and Garmire 2018). We most recently updated this method with an ensemble approach to further increase the stability (Poirion et al. 2019). Applying DeepProg on 32 cancer datasets from TCGA, we found that most cancers have two optimal survival subtypes, similar to the finding in HCC before. DeepProg yielded significantly (p-value=7.9e-7) better patient survival risk stratification than another multi-omics data integration Similarity Network Fusion (SNF). It also had excellent predictive accuracy in external validation cohorts, exemplified by two liver cancer (C-index 0.73 and 0.80) and five breast cancer datasets (C-index 0.68-0.73). Pan-cancer analysis showed that patients with the poorest survival subtypes were enriched in genes functioning in extracellular matrix modeling, immune deregulation, and mitosis. Therefore, it is possible to not only construct prognosis models using multi-omics data, but also learn important biological mechanisms that govern the prognosis.

## DISCUSSION

Patient survival prediction is clinical significant. New analytical approaches, such as machine-learning and deep-learning methods, will continue to drive forward the methodological improvement on multi-omics integration(S. Huang, Chaudhary, and Garmire 2017). To make "black-box" neural network models more interpretable and useful clinically, it is possible to combine prior biological knowledge with NN models. Patient survival risk stratification, rather than regression on survival time, is a practical consideration. We are currently comparing Cox-nnet (survival based) and DeepProg (classification based) approaches. Models that take into clinical phenotype, eg. individual subtype of cancer, can be more relevant (Chen et al. 2020). However, one must balance the practical issue of reduced sample size, in order to build such models, which may in turn have reduced predictability (esp. when applying deeplearning techniques). A good alternative may be including the patient clinical information as part of the model construction step, rather than sub-setting the patients. Recent development in single cell technologies, especially in single-cell multi-omics technologies, provokes us to think deeper the heterogeneity even within the same patient samples (Ortega et al. 2017; Zhu et al., n.d.). Efforts devoted to deconvolute the heterogenous signatures from tumor and other cell types, such as immune cells (Q. Huang et al., n.d.), will hopefully improve the quality of input data for multi-omics integration at the population scale. Besides the accuracy of the model, other factors need to be considered, such as efficiency, adaptability, biological insights, and actionability. A good prognosis model ideally will not only predict patient survival accurately, but also provide insights that suggest personalized therapeutic treatments.

## ACKNOWLEDGMENTS
We would like to thank the support by grants K01ES025434 awarded by NIEHS through funds provided by the trans-NIH Big Data to Knowledge (BD2K) initiative (www.bd2k.nih.gov), R01 LM012373 and R01 LM012907 awarded by NLM, and R01 HD084633 awarded by NICHD.

## CONFLICT OF INTEREST

The author declares no conflict of interest.

## FUNDING STATEMENT

This work was supported by grants K01ES025434 awarded by NIEHS through funds provided by the trans-NIH Big Data to Knowledge (BD2K) initiative (http://datascience.nih.gov/bd2k), P20 COBRE GM103457 awarded by NIH/NIGMS, and R01 LM012373 and R01 LM012907 awarded by NLM, R01 HD084633 awarded by NICHD to LX Garmire.